%% file: TOP2017_taejeong.tex
\documentclass[12pt]{article}
\usepackage{graphicx}

\newcommand\pubnumber{SNSN-323-63}
\newcommand\pubdate{\today}

\def\institute{Hanyang University, Department of Physics\\
Wangsimniro 222, Seoul 04763, Korea}

\def\Title#1{\begin{center} {\Large #1 } \end{center}}
\def\Author#1{\begin{center}{ \sc #1} \end{center}}
\def\Address#1{\begin{center}{ \it #1} \end{center}}

\newcommand\pubblock{\rightline{\begin{tabular}{l} \pubnumber\\
         \pubdate  \end{tabular}}}
\newenvironment{Abstract}{\begin{quotation}  }{\end{quotation}}
\newenvironment{Presented}{\begin{quotation} \begin{center} 
             PRESENTED AT\end{center}\bigskip 
      \begin{center}\begin{large}}{\end{large}\end{center} \end{quotation}}

\input econfmacros.tex
\begin{document}
\begin{titlepage}
\pubblock

\vfill
\Title{Differential cross sections of global event variables of $t\bar{t}$}
\vfill
\Author{Tae Jeong Kim on behalf of the ATLAS and CMS collaborations}
\Address{\institute}
\vfill
\begin{Abstract}
During the Run 2 period of the LHC, 
the ATLAS and CMS experiments have accumulated 
proton-proton collision data corresponding to 
an integrated luminosity of around 3 fb$^{-1}$ in 2015 and 36 fb$^{-1}$ in 2016
at a center of mass energy of 13 TeV.  
In the journey of finding new physics, it is of importance to understand the standard model which seems to be complete after the Higgs discovery.
Precision tests must be performed 
in every corner of the phase space since new physics can appear in any different places. 
In this proceedings, measurements of the differential cross sections of global event variables from the top quark pair productions at the both experiments 
using the data collected during the Run 2 period by the time of the year 2016 are presented.

\end{Abstract}
\vfill
\begin{Presented}
$10^{th}$ International Workshop on Top Quark Physics\\
Braga, Portugal,  September 17--22, 2017
\end{Presented}
\vfill
\end{titlepage}
\def\thefootnote{\fnsymbol{footnote}}
\setcounter{footnote}{0}

\section{Introduction}

The ATLAS~\cite{ATLASdet} and CMS experiments~\cite{CMSdet} at the CERN LHC
have collected proton-proton collision data corresponding to an integrated luminosity of 
around 3 fb$^{-1}$ in 2015 and 36 fb$^{-1}$ in 2016 
during the Run 2 period at the center of mass energy of 13 TeV.  
In the journey of finding new physics, it is of importance to understand the standard model 
which seems to be complete after the Higgs discovery.
Precision test must be performed in every corner of the phase space 
since new physics can appear in any different places.
In particular, the $t\bar{t}$ process is the main background for many searches for beyond the standard models.
Measurements with respect to global event variables not requiring kinematic reconstruction
such as jet multiplicity and missing transverse momentum (MET) from the top quark pair productions
are complementary. These global event variables are sensitive to dark matter searches.
Measurements with additional jets can allow us to check the validity of the QCD calculation involving a top quark pair 
plus additional quarks or gluons by comparing higher-order predictions against parton shower models and will lead us to better
understanding of systematic uncertainties on the theory modelling.   

In this proceedings, measurements of the differential cross sections of global event variables and also 
the variables of $\Delta R = \sqrt{ {(\Delta \phi)}^2 + {(\Delta \eta)}^2 } $
and invariant mass of the two additional jets are presented based on data collected the Run 2 period by the ATLAS and CMS experiments. 

\section{MC samples and particle-level objects}
In the ATLAS and CMS experiments, the Monte Carlo (MC) simulated samples for the $t\bar{t}$ signal are
generated using the POWHEG (v2) event generator at next-to-leading-order (NLO), 
interfaced with PYTHIA8 to provide the showering of the partons and to match soft radiation with the contributions from the matrix elements. 
For cross-checks and studies of systematic uncertainties, 
various event generators are used and described in Table~\ref{tab:mc}.
Note that not all of them are used to evaluate systematic uncertainties. 
\begin{table}[b]
\begin{center}
\begin{tabular}{|l|l|}  
\hline
Event generator          & Parton shower  \\ \hline
POWHEG (v2)              & PYTHIA8 (default) / HERWIG++ / HERWIG7 \\
MG5\_aMC@NLO (NLO) & PYTHIA8 with FxFx / HERWIG++ / HERWIG7 \\ 
MG5 (LO)           & PYTHIA8 with MLM  \\
SHERPA                   & default SHERPA tune  \\\hline
\end{tabular}
\caption{The Monte Carlo simulated samples used for the differential cross section measurements at the ATLAS and CMS experiments. Both experiments use the POHWEG event generator as a default $t\bar{t}$ simulated sample and SHERPA as cross checks. 
The different jet merging schemes for MG5\_aMC@NLO are used in the CMS experiment.}
\label{tab:mc}
\end{center}
\end{table}
%%%%%%%%%%%%%%%%%%%%%%%%%%%%%%%%%%%%%%%%%%%%%%%%%%%%%%%%%%%%%%%%%%%%%%%%%%%

Differential cross sections for the $t\bar{t}$ production are measured not only at parton level but also at particle level. 
One advantage of measuring the cross section at particle level is to 
decrease the uncertainty which can arise from extrapolating to unmeasurable phase space.
The particle-level jets are reconstructed by clustering all stable particle except 
the selected $e$, $\mu$ and radiated $\gamma$ as well as neutrinos (but do include those from hadron decay) using the anti-$k_{\rm T}$ with a parameter r = 0.4. 
To identify b-jets unambiguously, so called ``ghost matching" is used
by scaling down the b hadron momentum to a negligible value and including in the jet clustering. The b-jets are then identified by the presence of the corresponding ``ghost'' hadrons among the jet constituents. 
The MET in Section~\ref{sec:global} is defined as the vectorial sum of the transverse momenta of all neutrinos in the events, regardless of origin.  

\section{Jet multiplicity and additional jets}

\begin{figure}[htb]
\centering
\includegraphics[height=2.7in]{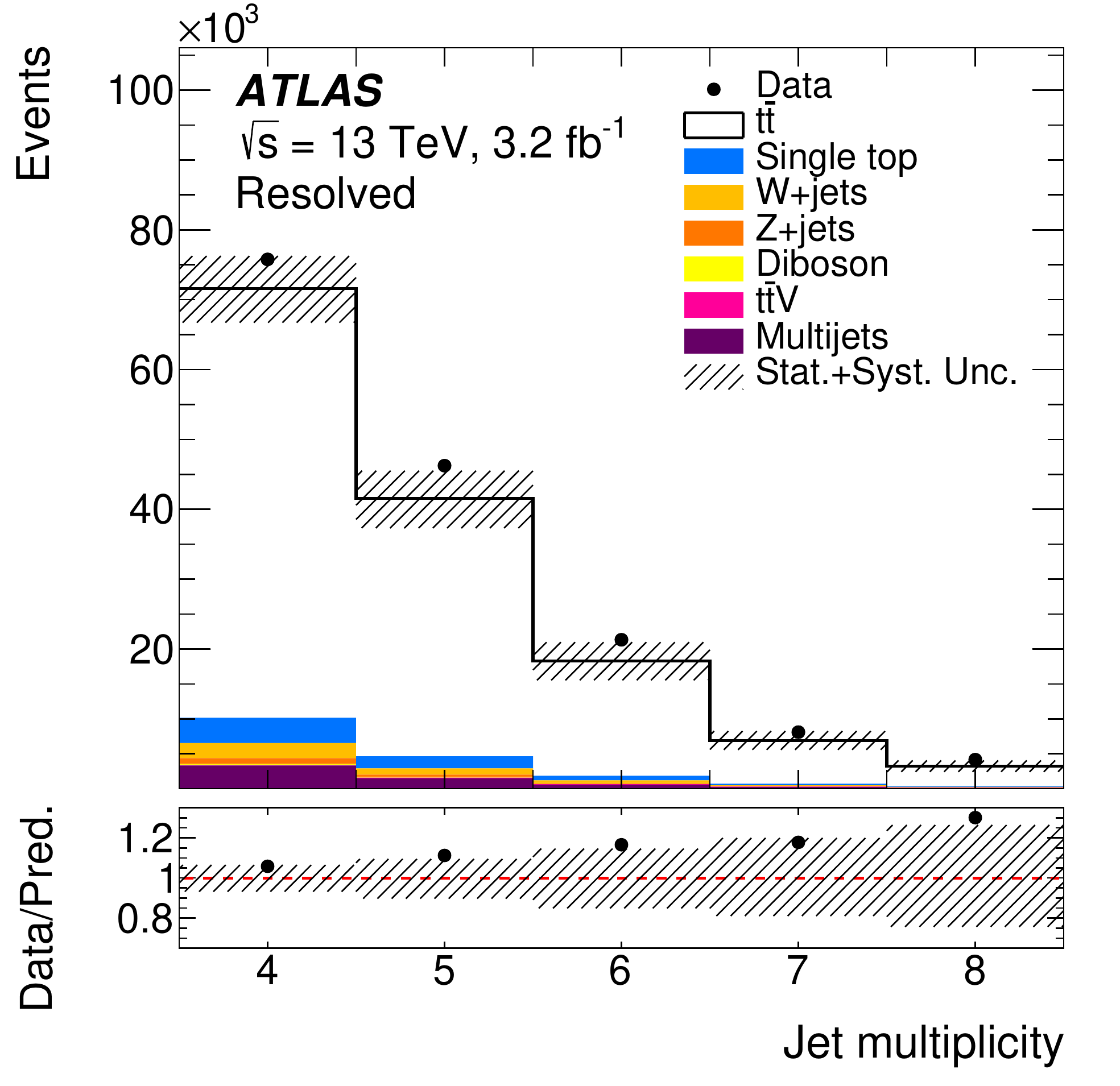}
\includegraphics[height=2.7in]{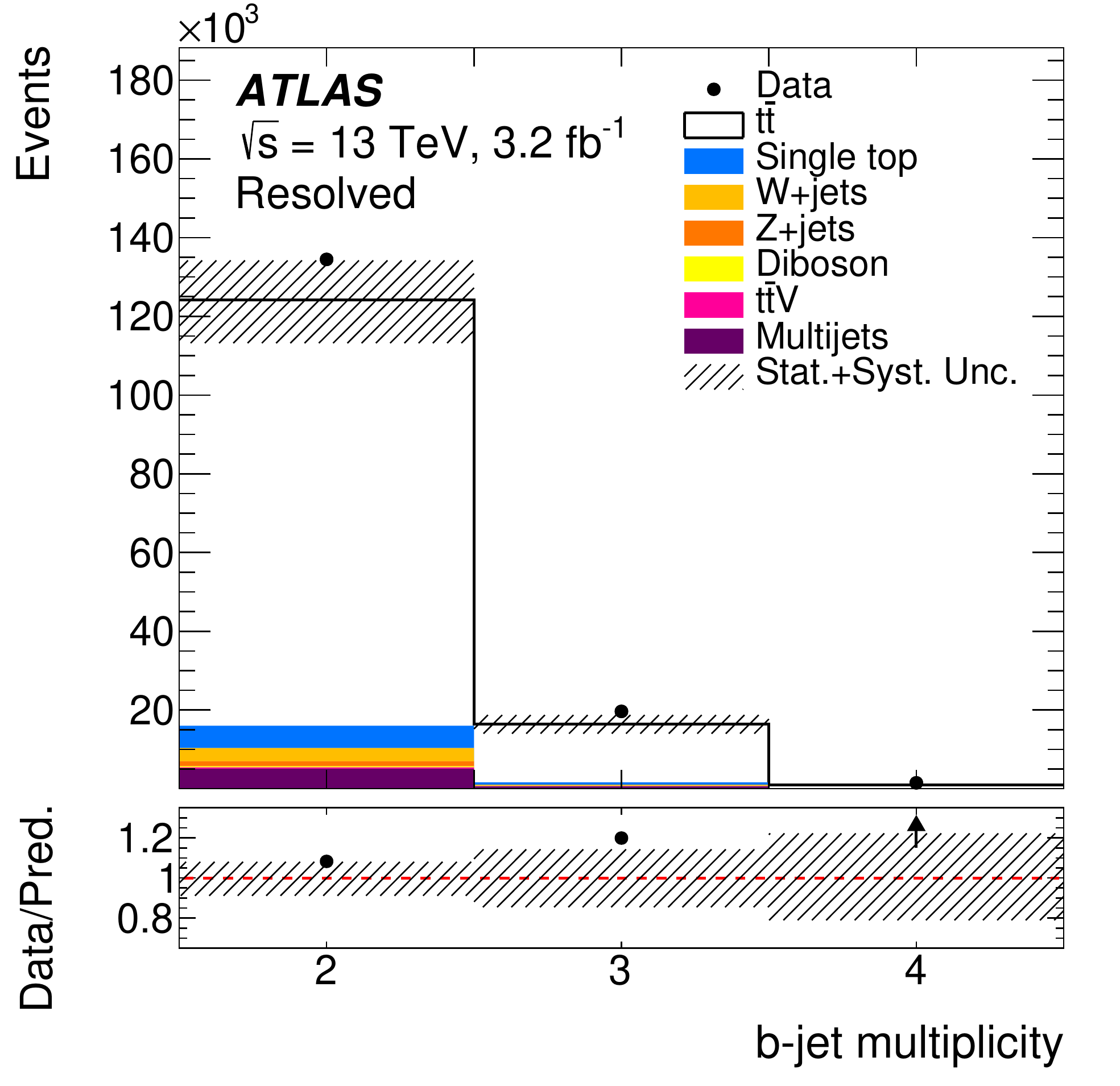}
\caption{Jet and b-jet multiplicity at the ATLAS experiment~\cite{atlasDiffLepJets}}
\label{fig:fig1}
\end{figure}

\begin{figure}[htb]
\centering
\includegraphics[height=2.7in]{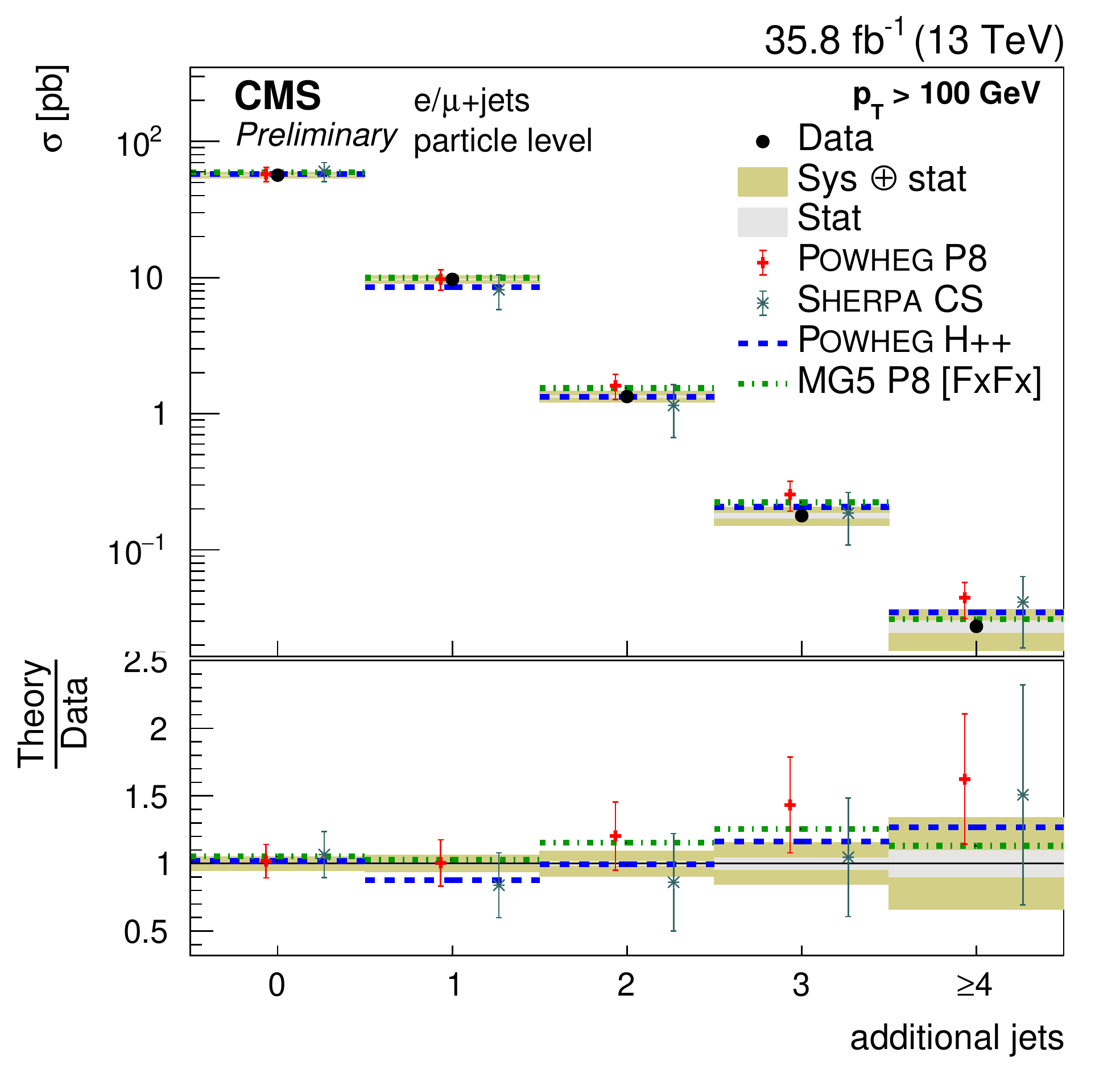}
\includegraphics[height=2.7in]{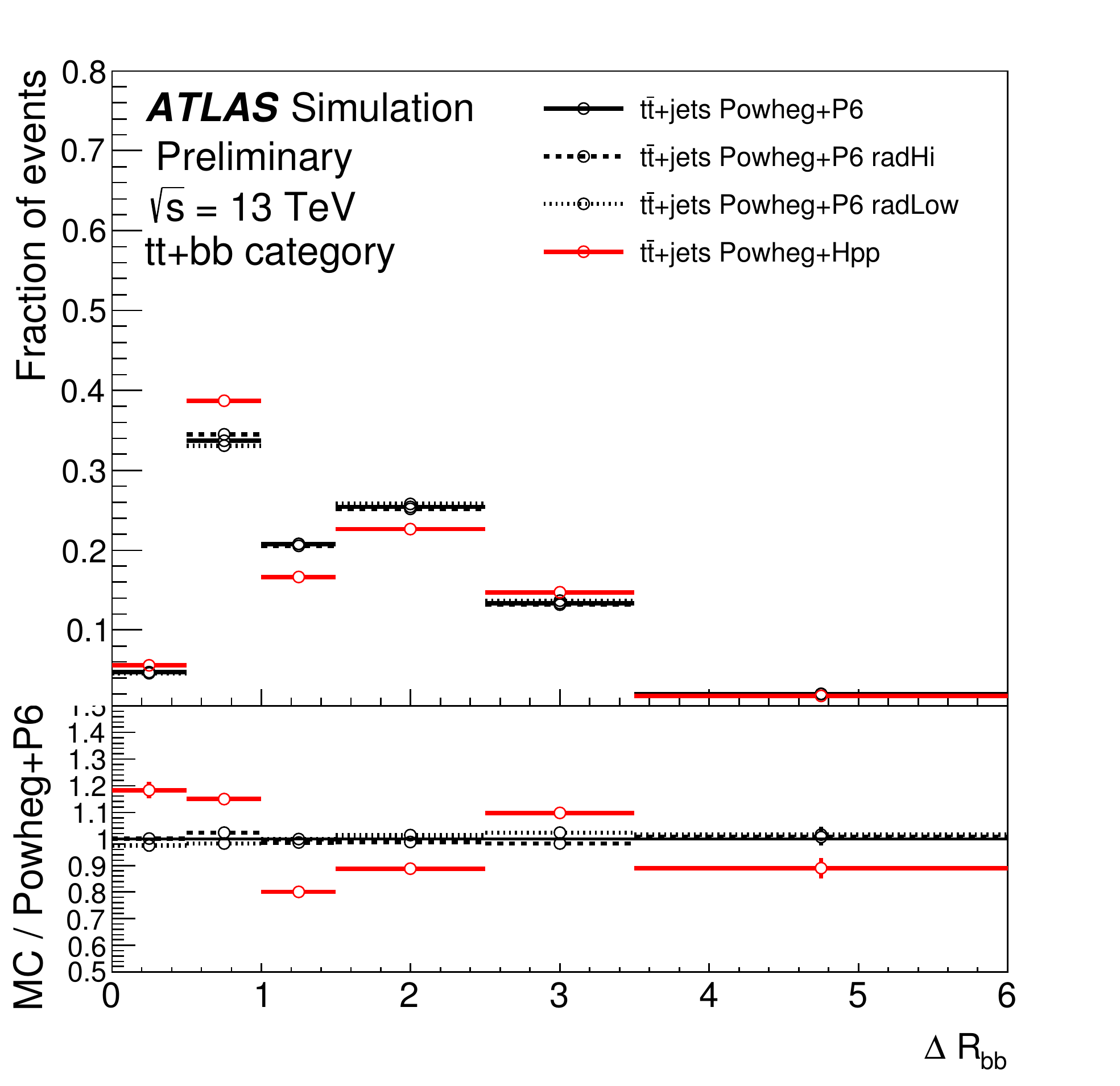}
\caption{Additional jet (left)~\cite{CMS-PAS-TOP-17-002} and $\Delta R$ between two additional b-jets distribution (right)~\cite{ATLttbb}.}
\label{fig:fig2}
\end{figure}

Jet and b-jet multiplicities in the $t\bar{t}$ process 
are measured using data corresponding to an integrated luminosity of 3.2 fb$^{-1}$ and 2.3 fb$^{-1}$  
at the ATLAS~\cite{atlasDiffLepJets}
and CMS experiment~\cite{CMS-PAS-TOP-16-021}, respectively.
The comparisons between data and MC simulations are shown in Fig.~\ref{fig:fig1}. 
Overall, the prediction has a tendency to underestimate data. 
In this measurement, the largest uncertainty comes from the jet energy scale/resolution and flavor tagging.  
The multiplicity for additional jets not from the $t\bar{t}$ system have also been measured
at the ATLAS experiment with 3.2 fb$^{-1}$~\cite{atlasAddLepJets} and at the CMS experiment with 36 fb$^{-1}$~\cite{CMS-PAS-TOP-17-002}. 
Figure~\ref{fig:fig2} (left) shows that
the multiplicity of additional jets is reasonably modelled by the MC simulation of the POWHEG event generator with PYTHIA8. The best agreement with data in this case comes from the MG5\_aMC@NLO with PYTHIA8. 
The simulated samples in this measurement are tuned for $h_{damp}$ and $\alpha_{s}^{ISR}$ 
using the results of the differential cross section at 8 TeV~\cite{CMS-PAS-TOP-16-021}. 

In the ATLAS collaboration, the study of the differential cross section of the anglur distribution between
two additional b-jets has been performed~\cite{ATLttbb}. 
Figure~\ref{fig:fig2} (right) shows the $\Delta R$ between the two additional b-jets together with
various NLO predictions with 4 flavor scheme (massive b quarks) indicated by the black-color line. 
A measurement of the differential cross section of the same variable in the $t\bar{t}b\bar{b}$ process at 8 TeV from the CMS experiment 
is also available in the Ref.~\cite{CMSttbb}.  
Both measurements show that the parton shower event generator HERWIG++ shows some deviations
in the $t\bar{t}b\bar{b}$ process.  

Highly boosted top quark in the all hadronic channel is also reconstructed 
with the threshold of $p_{\rm T}$ $>$ 500 GeV and 350 GeV for a leading 
and second leading top quark, respectively.
The scalar sum of the transverse momenta of the two top-quark jets is compared together 
with the distributions created with the POWHEG and aMC@NLO event generators
interfaced to the parton shower models of PYTHIA6 or HERWIG++ in Ref.~\cite{ATLASboosted}.

\section{Global event variables}
\label{sec:global}
\begin{figure}[htb]
\centering
\includegraphics[height=2.7in]{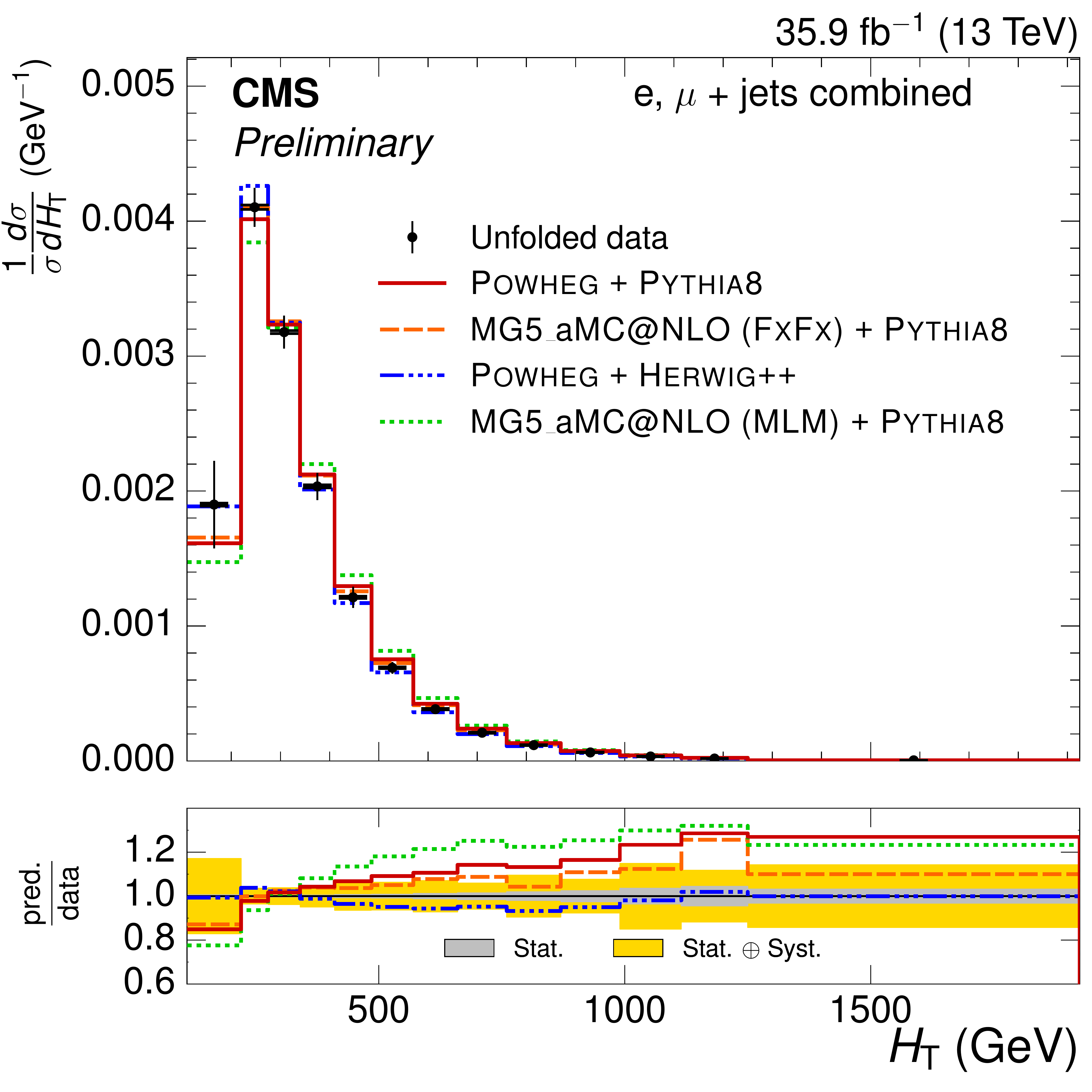}
\includegraphics[height=2.7in]{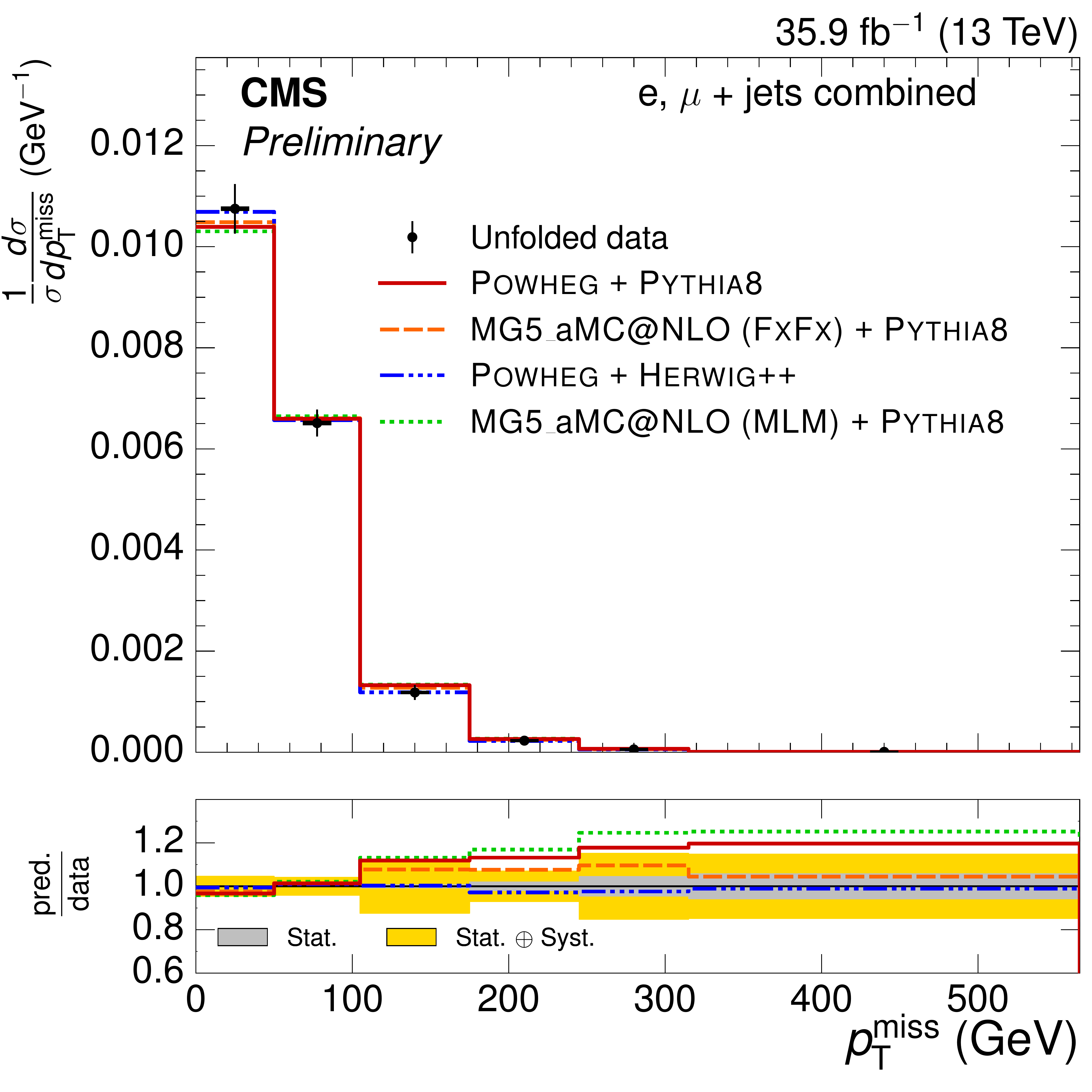}
\caption{$H_{\rm T}$ (left) and $P^{\rm miss}_{\rm T}$ (right) distributions at the CMS experiment~\cite{CMS-PAS-TOP-16-014}.}
\label{fig:fig3}
\end{figure}

In the CMS experiment, the differential cross sections of global event variables 
that do not require the reconstruction of the $t\bar{t}$ system are measured 
based on the full 2016 data corresponding to an integrated luminosity of 36 fb$^{-1}$~\cite{CMS-PAS-TOP-16-014}.   
Measured variables are the MET ($p_{\rm T}^{\rm miss}$), 
the scalar sum of the transverse jet momenta ($H_{\rm T}$), the transverse momentum of 
the leptonically decaying boson ($p_{\rm T}^{\rm W}$), the scalar sum of all transverse momenta of the particles ($S_{T}$), 
the jet multiplicity ($N_{jets}$), and the transverse momentum of the lepton ($p_{\rm T}^{l}$).
Figures~\ref{fig:fig3} show the $H_{\rm T}$ (left) and $P^{\rm miss}_{\rm T}$ (right) distributions.
The MG5\_aMC@NLO (leading-order) simulation does not accurately describe distributions in data while three other predictions have general agreement.
Using these 6 global variable distributions, a global $\chi^2$ test 
between the absolute cross sections in data and several simulation models was performed
assuming there is no correlation between the variables. 
Without including uncertainties in the predictions, 
the POWHEG+HERWIG++ simulation has the best agreement in the studied distributions with $\chi^2$/ndf = 62.5/62.
The POWHEG+PYTHIA8 simulation has a general agreement with $\chi^2$/ndf = 100.4/62.

\section{Conclusion}
Measurements of the differential cross section for jet multiplicity and additional jets as well as 
the global event variables from the $t\bar{t}$ are 
presented by comparing with various MC predictions. 
It has been shown that generally the event generators of 
POWHEG with PYTHIA8, POWHEG with HERWIG++, MG5\_aMC@NLO (FxFx) have 
a general agreement with data.
When it comes to the differential cross section of the $\Delta R$ of the additional b-jets, the parton shower event generator HERWIG++ shows some deviations.
There is no single simulation that can provide a good description 
of all variables simultaneously.
In the ATLAS and CMS experiments, many differential measurements at 13 TeV with high precision 
are available and 
need to be compared with each other to improve the theory modelling.
More differential cross section measurements with the full data of 36 fb$^{-1}$ are expected to come soon.

\end{document}

%% file: econfmacros.tex
\def\beq{\begin{equation}}
\def\eeq#1{\label{#1}\end{equation}}
\def\eeqn{\end{equation}}

\def\beqa{\begin{eqnarray}}
\def\eeqa#1{\label{#1}\end{eqnarray}}
\def\eeqan{\end{eqnarray}}

\let\bar=\overbar

\def\Dslash{\not{\hbox{\kern-4pt $D$}}}
\def\dslash{\not{\hbox{\kern-2pt $\del$}}}

\def\msb{{\bar{\ssstyle M \kern -1pt S}}}